\newif\ifsubmode
\submodetrue

\ifsubmode
  \documentclass[12pt,preprint]{aastex}
\else

\fi

\usepackage{graphicx}
\newcommand{\lta}{\lesssim}                                               
\newcommand{\gta}{\gtrsim}                                                
\newcommand{\kms}{\,\rm km\,s^{-1}}                                       

\newenvironment{inlinefigure}{
\def\@captype{figure}
\noindent\begin{minipage}{0.999\linewidth}\begin{center}}
{\end{center}\end{minipage}\smallskip}

\shorttitle{The $z \sim 4$ Lyman Break Galaxies: Colors and Theoretical Predictions}
\shortauthors{Idzi et al.}

\received{2003 May 24}

\begin{document}

\title{The $z \sim 4$ Lyman Break Galaxies: Colors and Theoretical Predictions\altaffilmark{5,6}}

\author{
Rafal Idzi\altaffilmark{2},
Rachel Somerville\altaffilmark{1},
Casey Papovich\altaffilmark{4}, 
Henry C. Ferguson\altaffilmark{1,2}, 
Mauro Giavalisco\altaffilmark{1},
Claudia Kretchmer\altaffilmark{2}, 
Jennifer Lotz\altaffilmark{3}
}

\altaffiltext{1}{Space Telescope Science Institute, 3700 San Martin Drive, 
Baltimore, MD  21218, USA; somerville@stsci.edu, ferguson@stsci.edu, mauro@stsci.edu}

\altaffiltext{2}{Department of Physics and Astronomy, The Johns Hopkins
University, 3400 N. Charles St., Baltimore, MD 21218; idzi@stsci.edu, claudia@pha.jhu.edu}

\altaffiltext{3}{Department of Astronomy and Astrophysics, University
of California, Santa Cruz, CA 95064; jlotz@scipp.ucsc.edu}

\altaffiltext{4}{Steward Observatory, The University of Arizona, 933 North
 Charry Avenue, Tuscon, AZ 85721; papovich@as.arizona.edu}
 
\altaffiltext{5}{Based on observations taken with the NASA/ESA Hubble
Space Telescope, which is operated by the Association of Universities
for Research in Astronomy, Inc.\ (AURA) under NASA contract
NAS5--26555}

\altaffiltext{6}{Based on observations collected at the European Southern Observatory,
 Chile (ESO Programmes 168.A-0485, 64.0-0643, 66.A-0572, 68.A-0544)}
 
 \begin{abstract}
We investigate several fundamental properties of $z \sim 4$ Lyman-break 
galaxies by comparing observations with the predictions of a
semi-analytic model based on the Cold Dark Matter theory of
hierarchical structure formation. We use a sample of $B_{435}$-dropouts 
from the Great Observatories Origins Deep Survey, and complement the ACS 
optical $B_{435}$, $V_{606}$, $i_{775}$, and $z_{850}$ data with the 
VLT ISAAC $J$, $H$, and $K_{s}$ observations. We extract $B_{435}$-dropouts 
from our semi-analytic mock catalog using the same color criteria and magnitude limits
that were applied to the observed sample. We find that the $i_{775} - K_{s}$ 
colors of the model-derived and observed $B_{435}$-dropouts are in good agreement. 
However, we find that the $i_{775}-z_{850}$ colors differ significantly,
indicating perhaps that either too little dust or an incorrect extinction curve have been used. 
Motivated by the reasonably good agreement between the model and observed data
we present predictions for the stellar masses, star formation rates, and ages for the $z \sim 4$ 
Lyman-break sample. We find that according to our model the color selection criteria used to 
select our $z \sim 4$ sample surveys $67 \%$ of all galaxies at this epoch down to $z_{850} < 26.5$. 
We find that our model predicts a $\sim 40 \%$ mass build-up between the $z \sim 4$ and $z \sim 3$ 
epochs for the UV rest-frame $L^{*}$ galaxies. Furthermore, according to our model, at least $50 \%$ 
of the total stellar mass resides in relatively massive UV-faint objects that fall below our observational 
detection limit. 
\end{abstract}
 
 \section{Introduction}
 The "Lyman-break'' color selection technique has been shown to be a
 highly effective means of selecting galaxies at high redshift $z \gta 2$ \citep{SAG1999,MFD1996}.
 Fairly large samples of Lyman-break galaxies (LBGs) in the redshift
 range $2 \lta z \lta 3.5$ have been compiled, and the characteristics
 of these objects have been studied with much vigor over the past
 several years. Optical and NIR photometric colors have been used to
 ascertain various attributes of these high-redshift objects. For
 example, colors spanning the rest-frame Balmer break ($\lambda \sim
 4000$ \AA) have been used to constrain the range of possible stellar
 ages and masses of these galaxies through application of maximum-likelihood 
 techniques to synthesized stellar population models \citep{PDF2001,SSA2001}. 

 As well, many researchers have compared theoretical models of galaxy
 evolution with various aspects of the observations \citep{WSB2001,BCF1998}.
 \citet[][hereafter SPF2001]{SPF2001} have used semi-analytic models to
 perform detailed comparative studies on the $z \sim 3$ sample. In SPF2001
 it was shown that the luminosity function and rest-frame UV and UV-optical colors of
 $z \sim 3$ LBGs could be reproduced well by the models, implying that
 the ages and dust contents of the model galaxies were consistent with
 the observations. 
 
 In this letter, we compare the colors of observed $B_{435}$-dropouts, selected from ACS imaging 
 from the Great Observatories Origins Deep Survey (GOODS) and supplemented with 
 deep NIR ($JHK_{s}$) data from ISAAC on the VLT, with the colors of model-derived $B_{435}$-dropouts. 
 We then present model predictions  for physical properties of these objects, such as stellar masses,
 ages, and star formation rates. We also investigate selection effects on the stellar mass content 
 for the redshift that we sample, and we look at the stellar mass build-up between the $z \sim 4$ 
 and $z \sim 3$ epochs.

 All quoted magnitudes throughout the letter are in the AB magnitude system \citep{O1974}.

 \section{Observations and Measurements}\label{observations}

 We use three epochs of HST ACS observations of the Chandra Deep Field
 South (CDF-S) obtained as part of the GOODS program. The data consist
 of image mosaics in all four ACS bands $B_{435}$, $V_{606}$,
 $i_{775}$, and $z_{850}$ (the ACS F435W, F606W, F775W, F850LP filters,
 respectively) spanning wavelength range of $0.4$ to 1 $\micron$, and
 covering approximately a field of view of $10 \arcmin$ x $16 \arcmin$. 
 The specifics of data acquisition and reduction, as well as object detection and
 photometry can be found in \citet{G2003a}.

 We supplement our ACS data set with the VLT ISAAC near-infrared imaging. 
 The NIR data consists of the J, H, and $K_{s}$ bands that extend our wavelength coverage 
 out to $\lambda \sim 4400$ \AA (rest-frame $z \sim 4$). We combine the ACS and ISAAC data 
 using a method that optimally matches the relative photometry between these bands 
 \citep{PapPhD,P2003}.  
 
 \section{Theoretical Model}\label{theoretical model}

 We make use of a semi-analytic model, based on the hierarchical
 structure formation paradigm in a $\Lambda$CDM cosmology
 ($\Omega_{\Lambda} = 0.7$, $\Omega_{m} = 0.3$, $\rm h = H_{0}/100$ 
 $\kms$$\rm Mpc^{-1} = 0.7$). The model treats the formation of structure
 via a hierarchical `merger tree', and includes a treatment of gas
 cooling, star formation, supernova feedback and chemical enrichment,
 galaxy mergers, stellar populations and dust (see \citet{SP1999} and
 SPF2001 for details). We follow halo merger histories down to a
 circular velocity of $\rm V_c =30 \kms$, as gas collapse and star
 formation in smaller halos is assumed to be suppressed by the presence
 of a photoionizing background \citep[see][and references
 therein]{S2002}. We use the multi-metallicity stellar SEDs from the
 STARDUST models \citep{DGS1999}, with a Kennicutt IMF. Dust extinction
 is modeled in a similar manner as in SPF2001. We use this model to
 produce a `mock-GOODS' catalog with the same geometry, sky area,
 filter passbands, etc. as the real GOODS.

 \section{Galaxy Sample}\label{galaxy sample}

 The $B_{435}$-dropout sample was chosen using the color selection criteria described in detail by \citet{G2003b}. 
 The sample was refined by applying detection limits at (S/N)$_{z,i} > 5$ for the ACS data set, and another one at
 (S/N)$_{K_{s}} > 1$ for the matched NIR set. These signal-to-noise restrictions limited our sample 
 to galaxies with $z_{850} < 26.5$, which helped curtail the presence of spurious detections. The relatively faint 
 $K_{s}$ cut-off was necessitated by the desire for a large enough sample for our study and did not significantly
 affect our conclusions. After further visual inspection, we chose 136 color-selected galaxies for analysis.

 Using the same color selection criteria that were applied to the observed sample we produced a model-derived
 $B_{435}$-dropout sample. We then applied a magnitude cut-off of $z < 26.5$ in order to comply with the
 observational detection limit. Before applying the color selection criteria we incorporated simulated observational scatter 
 into our model-derived photometric catalogs. The observational scatter was drawn from a Gaussian distribution with
 the typical signal-to-noise values found in the observed CDF-S data set. 

 Due to the dearth of spectroscopic data, empirical Monte Carlo simulations were performed to estimate 
 the redshift distribution of the observed $B_{435}$-dropout sample. These simulations
 were based on artificial LBGs distributed over a wide redshift range ($2.5 < z < 8$) with assumed distribution
 functions of UV luminosity, SED, morphology, and size, adjusted to match the colors of observed $B_{435}$-dropouts
 observed at $z \sim 4$. Reader is urged to see \citet{G2003b} for a detailed discussion on these simulations. 
 The redshift distribution of the simulated color-selected sample was found to have a mean value of  $z \sim 3.78$ with 
 a standard deviation of $\pm 0.34$. 
 
 One of the interesting tests that we can perform is to test for the incompleteness of our color-selected sample. 
 According to the Monte Carlo simulations, $73 \%$ of all the simulated $B_{435}$-dropout galaxies down to $z_{850} < 26.5$ 
 and in the interval $3.44 < z < 4.12$ are recovered using our color selection criteria.  When we apply the color criteria to 
 our model, we select $67 \%$ of all model galaxies in the same $3.44 < z < 4.12$ redshift range, down to the same 
 limiting magnitude of $z_{850} < 26.5$. This implies that the simulations performed by \citet{G2003b} and the 
 semi-analytic model show concordant incompleteness estimates with respect to the $B_{435}$-dropout selection
 technique in the above redshift range and down to our detection limit. 

 \section{Galaxy Properties}\label{galaxy properties} 

 The $i_{775} - K_{s}$ vs. $i_{775} - z_{850}$ (1550 - 4400 vs. 1550 - 1700 \AA, rest-frame $z \sim 4$)
 colors for the observed and model-derived $B_{435}$-dropout samples are presented in Figures 1a and 1b, 
 respectively. A Kolmogorov-Smirnov test reveals a $19 \%$ likelihood that the $i_{775} - K_{s}$ colors are 
 drawn from the same underlying distribution for the two $B_{435}$-dropout samples. A probability of 
 2 X $10^{-8}$ is obtained for the corresponding $i_{775} - z_{850}$ colors. This indicates a relatively good 
 agreement in the $i_{775} - K_{s}$ color distribution between the two samples, but a poor correlation in the 
 $i_{775} - z_{850}$ colors. 

 The $i_{775} - z_{850}$ colors probe the slope of the UV continuum, which is believed to be
 primarily an indicator of internal dust content  in young stellar populations \citep[e.g.][]{MHC1999}.
 In our model we used dust normalized against the $z = 0$ data and a Galactic extinction curve.
 The apparent disparity could be fixed by employing a different extinction curve \citep{C1997}, 
 or including the expected dependence of the extinction on the age of the stellar populaton \citep{CF2000}.

 Aside from the photometric analysis, we compared the luminosity function from our model with the one 
 for G-dropouts from \citet{SAG1999} and B-dropouts from HDF-N to ensure proper number counts. 
 We found good agreement, which is not surprising since our model is very similar to the one that was used 
 (and optimized) in SPF2001. In fact, the reader is urged to refer to SPF2001 for a detailed discussion on this 
 topic.
 
 Several researchers \citep[e.g.][]{PDF2001,SSA2001} have used simple
 parameterized star formation histories and stellar population
 synthesis models to estimate physical properties of LBG's at $z \sim 3$
 using the optical-NIR photometry as a constraint on the star formation
 history. Here, as we have shown that our model reproduces (modulo dust) 
 the observed color distribution for objects selected via their $B_{435} - V_{606}$ 
 and $V_{606} - i_{775}$ colors and corresponding magnitude limits, 
 we argue on similar grounds that our model should reproduce the statistical distribution 
 of underlying stellar ages and masses of the observed $B_{435}$-dropout population. 
 Thus we can use our models to obtain estimates for some of these quantities. 

 In Figure 2 we show the stellar masses of our color-selected model galaxies. 
 We note that the stellar masses range from $10^{8}$ to $10^{10}$ $\rm h^{-2}$ 
 $\rm M_{\sun}$, which is roughly two orders of magnitude less than the stellar masses of
 the present day $L_{*}$ spirals and ellipticals --- this indicates that, as with the $z \sim 3$ population, 
 the $z \sim 4$ Lyman-break galaxies are not the fully assembled progenitors of the present-day $L
 > L_{*}$ galaxies \citep{GSM96,S96a}, and that several generations of merging events must take 
 place between $z \sim4 $ and the present epoch. The median mass of $\rm Log[M_{star}] \sim 9.26$
 ($\rm M_{\sun}$ $\rm h^{-2}$) is 0.5 dex less than for the $z \sim 3$ galaxies studied 
 in SPF2001. This implies a stellar mass build-up between the two epochs. 
 To further explore the last point we have looked at the mean stellar masses of all $L^{*}$ 
 galaxies measured in rest-frame UV and predicted by our model at $z \sim 3$ and $z \sim 4$. 
 We find that at $z \sim 3$, with $m_{*} = 24.358$ (UV rest-frame), we get a mean value of
 $\rm Log[M_{star}] = 9.74$ ($\rm M_{\sun}$ $\rm h^{-2}$) in a $m_{*} \pm 0.5$ magnitude interval, 
 and at $z \sim 4$, with $m_{*} = 24.998$ (UV rest-frame), we get a mean value of 
 $\rm Log[M_{star}] = 9.58$ ($\rm M_{\sun}$ $\rm h^{-2}$), again in a $m_{*} \pm 0.5$ magnitude interval.
 This corresponds to a mass build-up of approximately $\sim 40 \%$. This type of mass build-up 
 between $z \sim 4$ and $z \sim 3$ is similar to that inferred from the \citet{P2003}
 comparison of volume-averaged SEDs of observed Lyman-break galaxies. 

 Our model also provides us with the star formation rates (SFR) and ages of our
 color-selected model galaxies. Our model results tell us that the values of these two quantities for the 
 $z \sim 4$ sample are very similar to the ones found for the $z \sim 3$ sample studied by SPF2001. 
 We observe a branch of actively star-forming galaxies, and a broader branch of
 galaxies with lower SFR, corresponding to bursting and fading galaxies. The star formation rates for 
 the brightest galaxies in the model approach 100 $\rm M_{\sun}$ $\rm yr^{-1}$, similar to what has
 been found for $z \sim 3$ Lyman-break galaxies studied by SPF2001. 

 We have also looked at the distribution of \emph{stellar-mass-weighted} mean
 stellar ages. We found that the distribution is very broad and skewed toward ages of
 less than 300 Myr, with a median age of 200 Myr, and a peak at about
 100 Myr. There is, however, a tail of objects with older stellar populations, reaching a mean 
 age of $\sim 1$ Gyr --- close to the age of the universe at that redshift in our adopted cosmology 
 ($\sim 1.5$ Gyr). 

 Drawing on the stellar mass distribution results (Figure 2) we plot in Figure 3 the mass distribution 
 of all (i.e. not just color-selected) galaxies from our mock catalog, limited to galaxies with
 $\rm Log[M_{star}] > 9.26$ ($\rm M_{\sun}$ $\rm h^{-2}$) (the median value from Figure 2) 
 and spanning the redshift range of  $3.44 < z < 4.12$. As was mentioned in \S\ref{galaxy sample}, we select 
 $67 \%$ of model galaxies in this redshift range with our color selection criteria, down to a limiting magnitude of 
 $z_{850} < 26.5$. This corresponds to $62 \%$ of the total stellar mass available in that redshift range, down
 to that magnitude limit. Hence, with our color criteria, we 'observe' the majority of the stellar mass that resides in galaxies 
 brighter than $z_{850} < 26.5$ and spanning $3.44 < z < 4.12$. Figure 2 indicates that $50 \%$ of our color-selected 
 objects lie above $\rm Log[M_{star}] > 9.26$ ($\rm M_{\sun}$ $\rm h^{-2}$), we find though that only $50 \%$ of all model 
 galaxies more massive than $\rm Log[M_{star}] > 9.26$ ($\rm M_{\sun}$ $\rm h^{-2}$) are brighter than $z_{850} < 26.5$. 
 So for model-derived objects residing in $3.44 < z < 4.12$ we only sample $50 \%$ of the available stellar mass down to
 $z_{850} < 26.5$. The rest of the stellar mass resides in relatively massive UV-faint galaxies. The color selection 
 incompleteness down to $z_{850} < 26.5$ and the mass contained in the UV-faint objects with magnitudes $z_{850} > 26.5$ 
 are two effects that conspire to severely limit the amount of mass selected with optical surveys \citep{F2003}. 
 Given how well our model predicts the observed colors (modulo dust), number counts, and other properties, this points to 
 a substantial deficit in optically-selected galaxies, and by extension, the total stellar mass. Since our model catalog was limited 
 to galaxies with $z_{850} < 28.0$ values, the 'unseen' mass fraction estimate should be taken as a lower limit.

 \section{Summary and Conclusions}\label{summary} 

 We performed a comparative analysis of the color-selected $B_{435}$-dropout samples taken from GOODS 
 observations and a variant of a $\Lambda$CDM-based semi-analytic model. We found that the color
 selection technique used to obtain the $B_{435}$-dropout sample does a relatively good job in selecting a 
 complete census of galaxies spanning $z \sim 3.78 \pm 0.34$ and down to $z_{850} < 26.5$. The color
 selection incompleteness limits are roughly the same when applied to our model and the empirical Monte Carlo
 simulations preformed by \citet{G2003b}. We select $67 \%$ of all galaxies and $62 \%$ of stellar mass in that redshift range,
 down to $z_{850} < 26.5$. This lends further credence to applicability of the $B_{435}$-dropout color selection 
 technique outlined in \citet{G2003b}, which was designed based on simpler, more empirical models of galaxy SEDs at 
 this epoch, and has not, as yet, been extensively verified with spectroscopic data.

 We found a relatively good agreement between the model-derived and observed CDF-S $B_{435}$-dropout 
 galaxy $i_{775} - K_{s}$ colors. The $i_{775} - z_{850}$ colors showed much less agreement, however, we attribute 
 the discrepancy to potentially inadequate dust recipe employed in the model. In general, we found that the stellar age, 
 mass, and the star formation properties of the $z \sim 4$ sample were similar to the $z \sim 3$ sample studied by 
 SPF2001. We found a model-derived $\sim 40 \%$ stellar mass build-up for the UV rest-frame $L^{*}$ galaxies 
 between the two epochs. In addition we found that if we look at all galaxies predicted by our model, spanning 
 $3.44 < z < 4.12$ redshift range, then at least $50 \%$ of the stellar mass contained in objects with 
 $\rm Log[M_{star}] > 9.26$ ($\rm M_{\sun}$ $\rm h^{-2}$) is missed due to the $z_{850} < 26.5$ observational limit. 
 If our model is correct, and it does show reasonable agreement with respect to colors (modulo dust), number counts, 
 and other physical quantities, this result has significant implications for the completeness of the optically-selected 
 surveys at this epoch. 

 \acknowledgments

 Support for this work was provided by NASA through grant GO09583.01-96A from the 
 Space Telescope Science Institute, which is operated by the Association of Universities
 for Research in Astronomy, Inc.\ (AURA), under NASA contract NAS5-26555.

 Support for this work, part of the {\it Space Infrared Telescope 
 Facility (SIRTF)} Legacy Science Program, was provided by NASA through 
 Contract Number 1224666 issued by the Jet Propulsion Laboratory, 
 California Institute of Technology under NASA contract 1407.

 %\bibliographystyle{apj}
 %\bibliography{apjmnemonic,bib}

\clearpage

\begin{figure*}
\resizebox{0.5\textwidth}{!}{\includegraphics{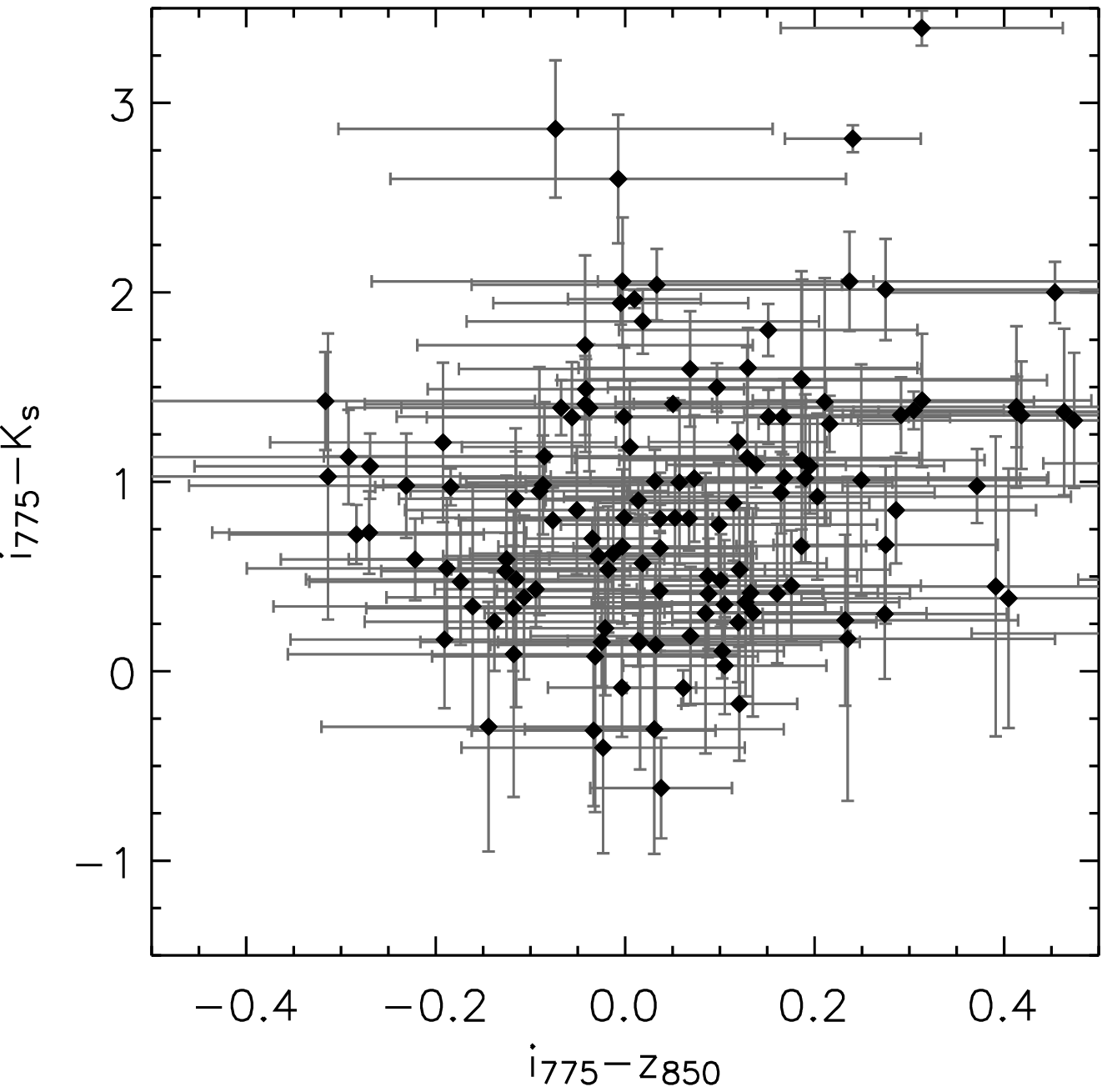}}
\resizebox{0.5\textwidth}{!}{\includegraphics{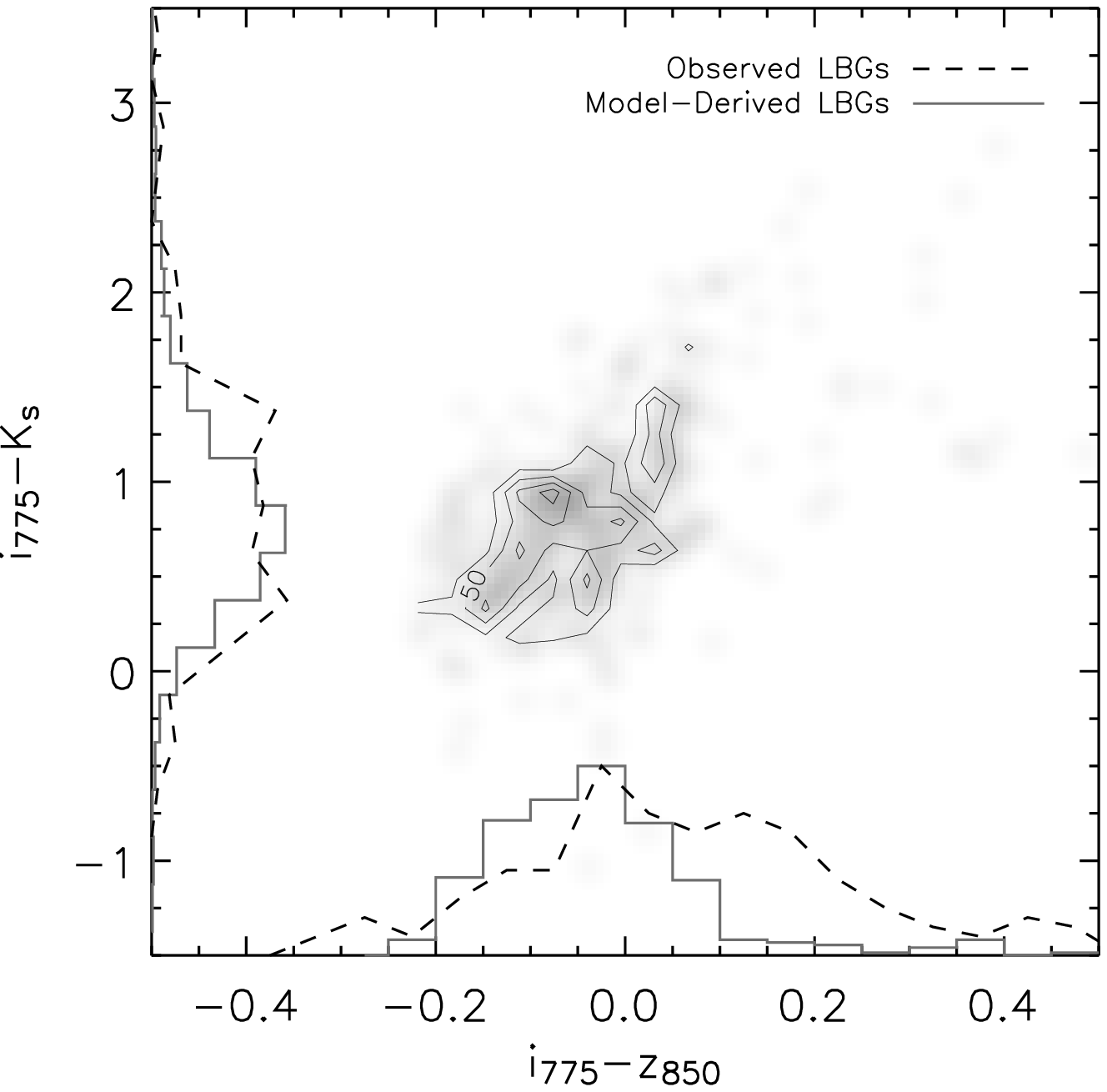}}
\figcaption{The $i_{775} - K_{s}$ versus $i_{775} - z_{850}$ colors of the $z \sim 4$ color-selected galaxies. 
Panel (a) shows the 136 observed $B_{435}$-dropout galaxies {\it filled-diamond} from the CDF-S. Panel (b) 
shows a Hess diagram of the `Collisional-Starburst' model galaxies with artificial observational scatter folded
in, and selected with the same color criteria and magnitude 
limits that were applied to the observed sample. Contours have been superimposed to help 
guide the eye; they represent the $30^{th}$, $50^{th}$, $70^{th}$, and $90^{th}$ percentiles. In addition, 
histograms are included for both the model ({\it gray solid-line}) and observed ({\it black dash-line}) galaxy colors. 
This figure illustrates the relative agreement in the $i_{775} -K_{s}$ color distributions for the model-derived and 
observed $B_{435}$-dropout galaxies, while at the same time illustrating an appreciable mismatch in the 
$i_{775} - z_{850}$ colors.\label{fig1}}
\end{figure*}
\clearpage

\begin{inlinefigure}
\resizebox{\textwidth}{!}{\includegraphics{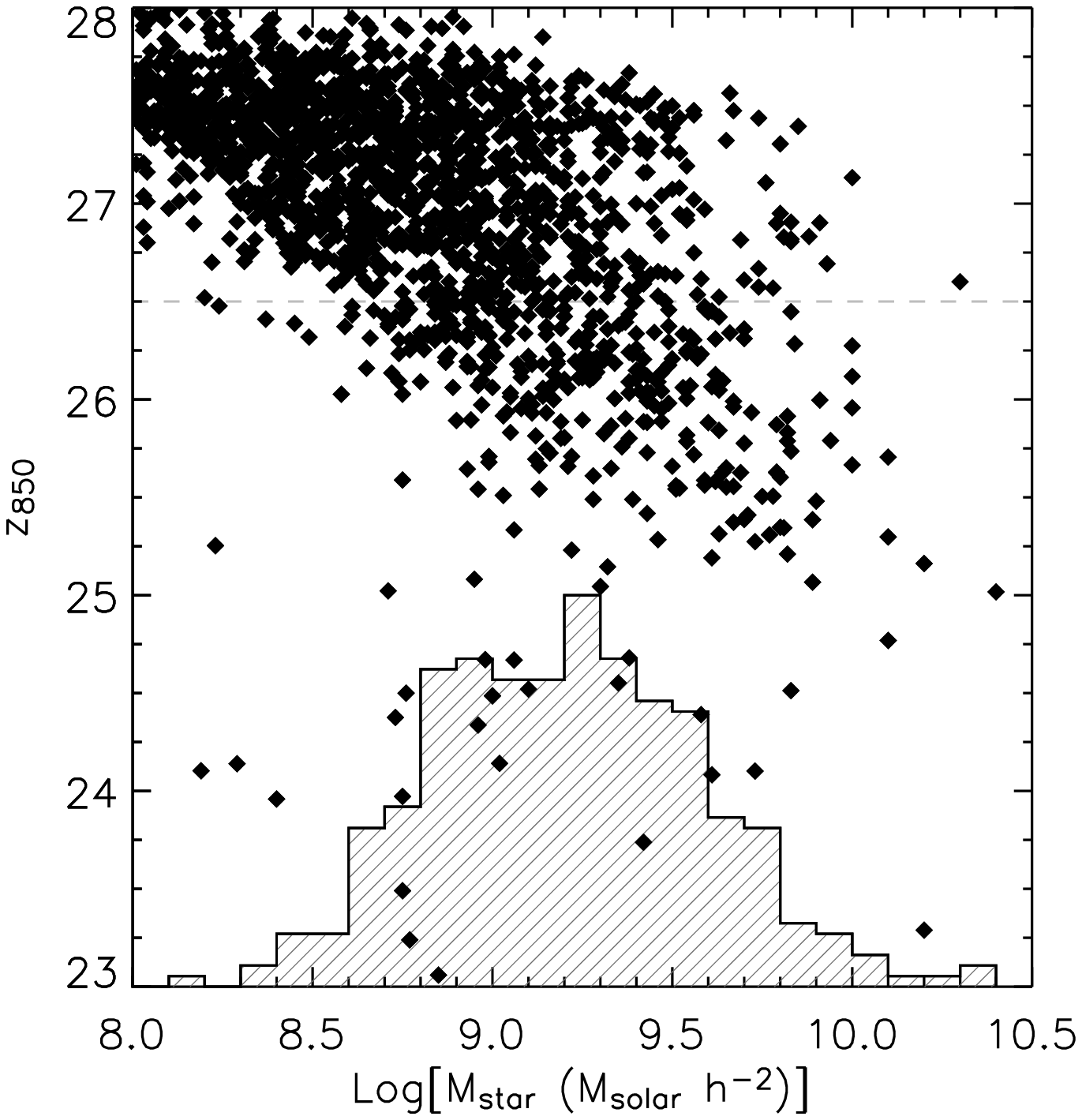}}
\figcaption{The stellar masses of the color-selected model galaxies. The {\it filled-diamond} symbols show stellar 
masses of the individual color-selected model galaxies vs. their corresponding $z_{850}$ magnitudes. 
The histogram shows the projected distributions for the same color-selected model galaxies 
with an imposed $z_{850} < 26.5$ magnitude limit (shown as a {\it gray dash-line}). The predicted masses are 
two magnitudes lower than the stellar masses of the present day $L_{*}$ spirals and ellipticals.\label{fig2}}
\end{inlinefigure}
\clearpage

\begin{inlinefigure}
\resizebox{\textwidth}{!}{\includegraphics{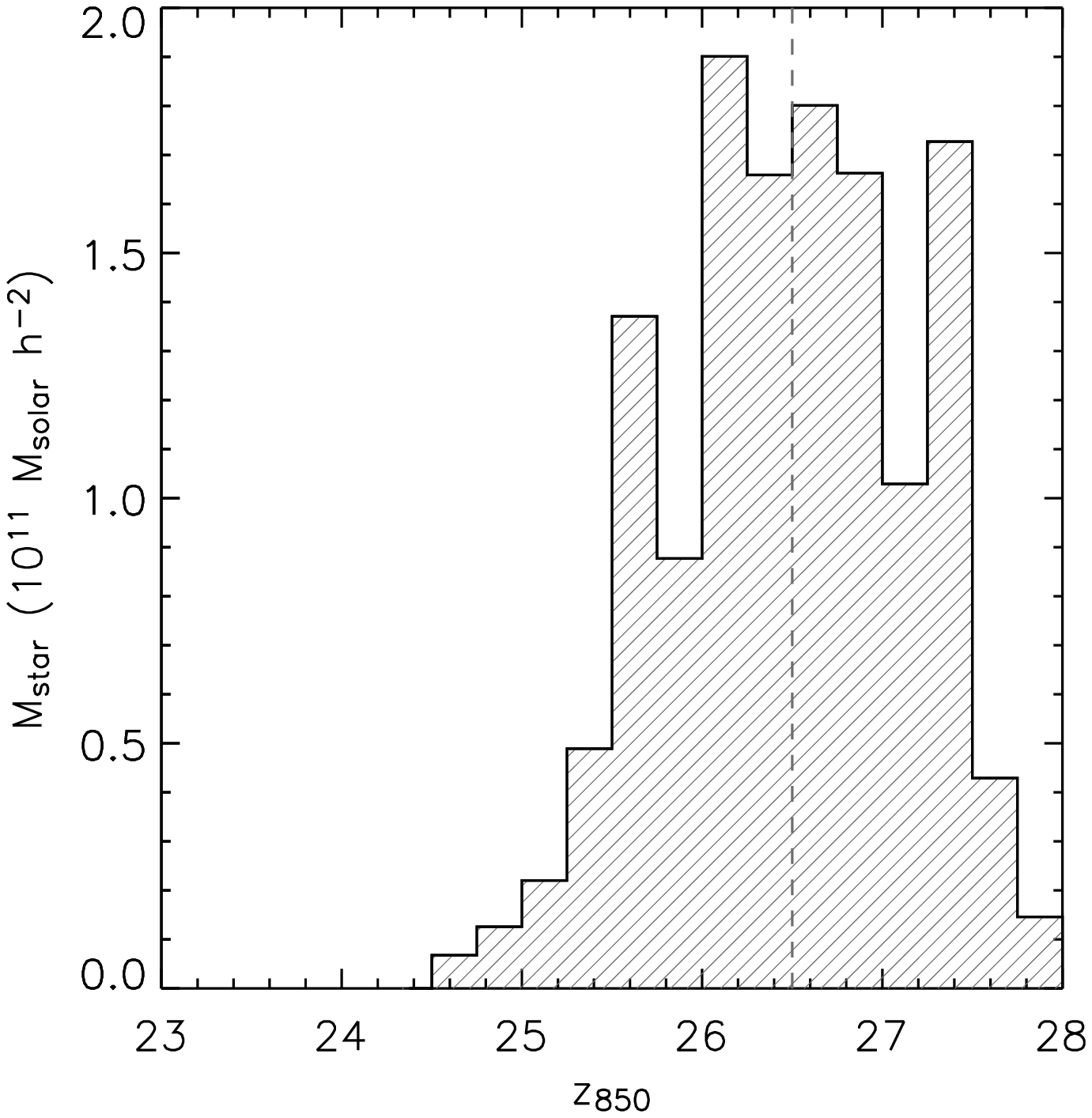}}
\figcaption{This figure illustrates the stellar mass distribution of all model-derived galaxies spanning the redshift 
range of $3.44 < z < 4.12$ and with $\rm Log[M_{star}] > 9.26$ ($\rm M_{\sun}$ $\rm h^{-2}$)
(the median value from Figure 2). Galaxies were binned into 0.25 magnitude intervals and weighted by their 
corresponding stellar mass. A {\it gray dash-line} is included to delineate our observational magnitude limit. This 
figure shows the amount of model-predicted mass potentially missed by the current optical surveys.\label{fig2}}
\end{inlinefigure}

\end{document}